\documentclass[aps,prl,twocolumn,nofootinbib,floatfix]{revtex4-1}

\usepackage{graphicx}  
\usepackage{bm}        
\usepackage{amssymb}   
\usepackage{amsmath}   
\usepackage{verbatim}
\begin{document}



\title{The finite size effects and the two-state paradigm of protein folding}


\author{Artem Badasyan} 
\email[Corresponding author's email:]{abadasyan@gmail.com}
\affiliation{Materials Research Laboratory \& School of Science, University of Nova Gorica, Vipavska 13, SI-5000 Nova Gorica, Slovenia}

\author{Matjaz Valant}
\affiliation{Materials Research Laboratory  \& School of Environmental Sciences, University of Nova Gorica, Vipavska 13, SI-5000 Nova Gorica, Slovenia}
\affiliation{University of Electronic Science and Technology of China, Institute of Fundamental and Frontier Sciences, Chengdu 610054, China}

\author{Joze Grdadolnik}
\affiliation{National Institute of Chemistry, Hajdrihova 19, SI-1000 Ljubljana, Slovenia}

\author{Vladimir N. Uversky}
\affiliation{Department of Molecular Medicine and USF Health Byrd Alzheimer's Research Institute, Morsani College of Medicine, University of South Florida,12901 Bruce B. Downs Blvd., MDC07, Tampa. Florida 33612, USA}



\date{\today}

\begin{abstract}
The coil to globule transition of the polypeptide chain is the physical 
phenomenon behind the folding of globular proteins. Globular proteins 
with a single domain usually consist of about 30 to 100 amino acid 
residues, and this finite size extends the transition interval of the 
coil-globule phase transition. Based on the pedantic derivation of the 
two-state model, we introduce the number of amino acid residues of a 
polypeptide chain as a parameter in the expressions for two 
cooperativity measures and reveal their physical significance. We 
conclude that the $k_2$ measure, related to the degeneracy of the 
denatured state, describes the number of cooperative units involved in 
the transition; additionally is found that the famous condition $k_2=1$ 
is just the necessary condition to classify the protein as the 
two-state folder. We also find that $\Omega_c$ is simply proportional 
to the square of the transition interval. This fact allows us to 
perform the classical size scaling analysis of the coil-globule phase 
transition. Moreover, these two measures are shown to describe different 
characteristics of protein folding. 

\end{abstract}

\pacs{}

\maketitle

\section{}

From the point of view of Polymer Physics the folding of a protein is similar to the coil-globule transition of a short polypeptide chain \cite{grosberg}. The coil-globule transition is known as the phase transition of first (in rigid) or second order (in flexible chains) \cite{grosberg,rmp1978}. By following the behaviour of the order parameter (degree of "nativeness")
$f_N(T) \in [1,0]$ or its counterpart ("denaturation" degree) 
$f_D(T)=1-f_N(T)$ it is possible to 
describe the phenomenon; the condition 
\begin{equation} \label{tf}
f_N(T_D)=f_D(T_D)=0.5
\end{equation}
defines the folding temperature $T_D$.

If there are no finite size effects or heterogeneity~\footnote{The 
account of heteropolymeric effects in the coil-globule transition is 
outside the scopes of the current study.}, the order parameter at the 
transition point undergoes an abrupt all-or-none fashion. Responsible 
for this coil-globule \textbf{phase transition} are strong correlations 
between repeat unit conformations, which occur due to the van der Waals 
interactions between the remote repeat units \cite{grosberg}. Changes in external conditions 
(temperature, pressure, pH, solution composition...) shift the equilibrium in these effective interactions from repulsion (good solvent regime) via neutral (ideal or theta conditions) to attraction (poor solvent regime), which forces the protein to fold. The hydrogen bonds, which are responsible for the formation of secondary structures, have a shorter span and influence the conformations locally. Therefore, according to the Landau-Pierls theorem  \cite{landau} the hydrogen bonds cannot \emph{per se} lead to coil-globule (phase) transition \cite{grosberg,hsc95}. However, in the presence of other long-range interactions, the formation of secondary structures can change the effective stiffness of the polypeptide chain, increase stability and thus promote the coil-globule transition. Indirect support for such a mechanism arises from the fact that both the coil-helix transition and protein folding occur at the same interval of external parameters \cite{uversky02}.

\textbf{Thermodynamic cooperativity} as a concept is often 
attributed to the sharpness of the phase transition, which results from 
the spatially correlated behaviour of the particles (in this case 
repeating units). The situation of the idealized first-order phase 
transition with correlations that extend throughout the system and lead 
to the discontinuity of order parameter corresponds to infinite 
cooperativity and the zero transition interval. When it comes to the 
folding of single domain globular proteins of just $N<100$  repeating 
units long, the limited system sizes impose constraints onto otherwise 
infinite correlations at transition point. Consequently, the 
folding happens over some small temperature interval 
$\Delta T (\neq 0)$, which needs to be estimated. Using the Taylor expansion 
cut at first order, it is possible to approximate the order parameter with 
the help of the tangent at transition point
\begin{equation} \label{apf}
f_D(T) \approx f_D^{appr}(T)=f_D(T_D)+f'_D(T)|_{T_D}(T-T_D).
\end{equation}
\noindent From the definitions of initial and final temperatures as $f_D^{appr}(T_1)=0$ 
and $f_D^{appr}(T_2)=1$, one can define the transition interval (see \emph{e.g.} 
\cite{frkamdeltat,moretal}) as 
\begin{equation} \label{dtdef}
T_2-T_1=\Delta T=f'_D(T)|_{T_D}^{-1}.
\end{equation}
\noindent The derivative of the order parameter at the transition point 
is the experimentally measurable quantity that provides access to 
information on the system's cooperativity. The temperature is not the 
only possible external parameter that can induce the transition. The experiments are 
often set by changing the concentration of denaturant such as urea or 
guanidinium chloride (GdmCl). After repeating the steps 
behind Eq.~\eqref{apf}, the resulting expression 
for the change in the number of bound denaturant molecules during the 
transition is
\begin{equation} \label{dndef}
n_2-n_1=\Delta n=f'_D(n)|_{n_D}^{-1},
\end{equation}
\noindent so that the thermodynamic cooperativity of transition can be still 
estimated by the measured slope of the transition curve at its middle point. 

\textbf{The two-state model} is the simplest among the folding models, 
yet very general and fruitful and therefore deserves a detailed, even pedantic derivation 
of its formulas. Within the two-state paradigm the presence of 
just two possible macroscopic states is assumed: the native globular 
state with the energy value $E_N$, and the denatured coil one with the 
energy $E_D$. To reflect the uniqueness of the native state, a 
degeneracy $g_N=1$ is attributed; a $g_D\gg 1$ degeneracy is set for 
the denatured state to reflect its greater conformational entropic 
freedom. Without any loss of generality one can assume $E_N=0$, $E_D \neq 0$ and write 
down the density of states for the two-state model
\begin{equation} \label{dos}
g(E)=\delta(E)+g_D\delta(E-E_D),
\end{equation}
\noindent where $\delta(x)$ is Dirac delta function, resulting in the 
partition function
\begin{multline} \label{pf}
Z(\beta)=\int_{0}^{\infty} dE \, g(E)  \,e^{-\beta E}=1+g_De^{-\beta 
E_D}= \\ [N]+[D],
\end{multline}
\noindent where $[...]$ is the number of repeat units in the native or 
denatured state, and $\beta=1/T$ is inverse temperature. The average 
energy is just the internal energy of the system, and follows directly 
as 
\begin{multline} \label{ie}
<E(\beta)>=\frac{\int_{0}^{\infty} dE \, g(E)  \,e^{-\beta E} 
E}{\int_{0}^{\infty} dE \, g(E)  \,e^{-\beta E}}=-\frac{d\log 
Z(\beta)}{d\beta}=\\ \frac{g_De^{-\beta 
E_D}}{1+g_De^{-\beta E_D}}E_D,
\end{multline}
\noindent leading to the heat capacity
\begin{multline} \label{cv}
C_V(\beta)=-\beta^2 \frac{d<E>}{d\beta}=(\beta E_D)^2 \frac{g_De^{-\beta 
E_D}}{(1+g_De^{-\beta E_D})^2}.
\end{multline}
\noindent 
The denaturation degree reads
\begin{multline}
f_D(\beta)=\frac{[D]}{[N]+[D]}=\frac{g_De^{-\beta E_D}}{1+g_De^{-\beta 
E_D}}=\\ \frac{<E(\beta)>}{E_D}=-\frac{1}{E_D}\frac{d\log 
Z(\beta)}{d\beta},
\label{dd}
\end{multline}
\noindent and the equilibrium constant
\begin{multline} \label{ec}
K_{eq}(\beta)=\frac{[D]}{[N]}=\frac{f_D(\beta)}{1-f_D(\beta)}=g_De^{-\beta E_D}.
\end{multline}
At transition point the numbers of repeat units in the native $N$ 
or the denatured $D$ state are equal and with the help of Eq.~\eqref{dd} we can express 
the transition temperature Eq.~\eqref{tf} and interval 
Eq.~\eqref{dtdef} in terms of two-state model parameters as
\begin{equation} \label{tfdt}
T_D=\frac{E_D}{\log g_D} \, ; \,\, \Delta T=\frac{4E_D}{\log^2 g_D}.
\end{equation}
Excluding $\log g_D$ from the last expression we re-derive the famous 
expression for the energetic price of transition between the two states 
as 
\begin{equation} \label{priv}
E_D=\frac{4 {T_D}^2}{\Delta T}.
\end{equation}
Privalov and Kheshinashvili \cite{priv74} refer to Eq.~\eqref{priv} as 
approximation, but as we have shown above, it is indeed exact within the two-state 
picture. Since all the above formulae are derived under the assumption 
of the existence of strictly two states, results can only be attributed to one 
\textbf{cooperative unit}, \emph{i.e.} a part of a molecule that 
undergoes the transition from $N$ to $D$ as a whole. Microcalorimetry 
allows the simultaneous measurement of the transition enthalpies for 
the whole protein molecule and for the cooperative unit \cite{priv79}. 
Potentiometric titration also allows the difference in the degree of 
ionization to be measured for the entire molecule and compared with the 
value for the cooperative unit \cite{her67}.

\textbf{The order} of a conformational transition can be evaluated by 
analysing the dependence of the slope of the transition on the 
molecular weight of the protein ($M$), which is linearly proportional to 
the degree of polymerisation $N$. It is clear that the slope of the phase 
transition in small systems depends on the dimensions of this system 
\cite{hill,grosberg}. In the case of first-order phase transition, the 
slope increases proportionally to the number of units in a system 
\cite{hill}, while the slope for second-order phase transition is 
proportional to the square root of this number \cite{grosberg}. 

\textbf{The system sizes} can be introduced by the 
reasonable assumption that each repeating unit of the polypeptide chain 
can be found in one out of $Q>2$ rotational isomeric states, only one 
of which corresponds to the native. Since there is $N$ such 
repeating units, the number of possible states in denatured conformation 
for the whole macromolecule and the additive energy of the system read
\begin{equation} \label{gen}
\begin{split}
g_D=(Q-1)^N ; \,\, E_D=\epsilon _D N.
\end{split}
\end{equation}
\noindent In view of Eq.~\eqref{tfdt} it 
means
\begin{equation} \label{tfdtN}
\begin{split}
T_D(N)=\frac{\epsilon_D}{\log (Q-1)} \,; \,
\Delta T(N)=\frac{4T_D}{\log(Q-1)}\frac{1}{N}.
\end{split}
\end{equation}
\noindent This is a very interesting result, which shows that within 
the two-state paradigm, denaturation temperature does not depend on 
system size. Instead, the transition interval is inversely proportional 
to $N$, which naturally leads to a zero interval at 
$N\rightarrow\infty$, just as it should in case of the phase transition.

This is a very interesting result, showing that within the 
two-state paradigm the denaturation temperature does not depend on system 
sizes. Instead, the transition interval is inversely proportional to $N$, 
naturally resulting in zero interval at $N\rightarrow\infty$, exactly as it 
should in case of the phase transition.

\textbf{The criterion of two-state cooperativity $k_2$} of protein folding has already been discussed in detail (see \emph{e.g.} \cite{kayachan2000,enzym2004} and references 
therein). It is defined as the ratio of 
vant Hoff and calorimetric enthalpy (energy)
\begin{equation} \label{k2}
k_2=\frac{\Delta E_{vH}}{\Delta E_{cal}},
\end{equation}
\noindent where the vant Hoff energy is
\begin{equation} \label{vhoffa}
\Delta E_{vH}=-\frac{d\log K_{eq}(\beta)}{d\beta}=E_D,
\end{equation}
and the amount of heat exchanged during the transition is calculated as 
the integral under the heat capacity curve:
\begin{equation} \label{calE}
\Delta E_{cal}=\int_{0}^{\infty} dT \, C_V(T)=E_D\frac{g_D}{1+g_D}.
\end{equation}
\noindent According to Eqs.~\eqref{gen},\eqref{vhoffa},\eqref{calE}, the resulting
\begin{equation} \label{k2fin}
k_2=\frac{\Delta E_{vH}}{\Delta E_{cal}}=1+1/g_D=1+O(e^{-N\log(Q-1)}),
\end{equation}
\noindent is an expression that asymptotically tends to $1$ (from 
above) for large $N$. It can be concluded that the two-state 
\emph{ansatz}, expressed in Eq.~\eqref{dos}, results in $k_2=1$, making 
it the \textbf{necessary} condition for the transition to be classified as a 
two-state. Please note, strictly speaking, it follows from nothing that $k_2=1$ 
means that the transition is two-state. In a certain sense, the 
condition is negative: if $k_2$ is different from unity, the transition 
cannot be two-state, while if it is close to the unity, it is not 
enough to conclude the two-state behaviour.

\textbf{The folding cooperativity measure} 
\begin{equation} \label{cmdef}
\Omega_c=\frac{T_D^2}{\Delta T}\frac{d f_D}{dT}\mid_{T=T_D}
\end{equation}
\noindent was proposed by Klimov and Thirumalai \cite{thirfolddes} to 
compare the cooperativities of different proteins. Based on a 
collection of experimental and simulation data of protein folding, a 
size scaling law for the folding cooperativity measure $\Omega_c \propto N^{1+\gamma}$ 
\cite{liprl} was later suggested, where $\gamma$ is a susceptibility exponent. 

\begin{figure}
\includegraphics[width=\columnwidth]{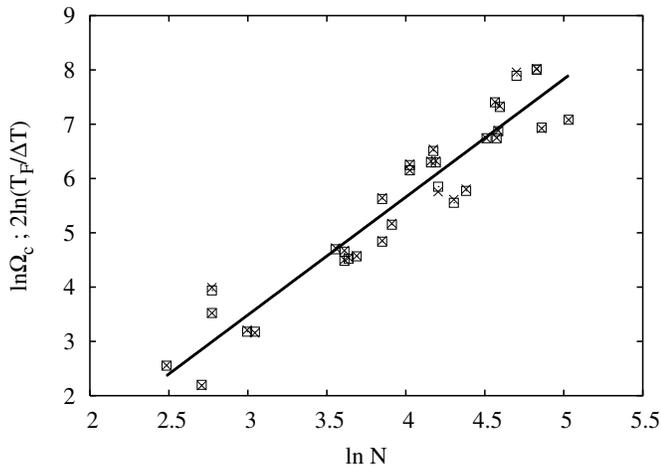}
\caption{\label{fig:epsart} The dependence of $\ln{\Omega_c}$ ($\times$) and $2\ln{({\frac{T_D}{\Delta T}})}$ ($\square$) vs $\ln{N}$.
Straight lines, corresponding to the linear fits for both data point collections are indistinguishable on the graph.}
\end{figure}

Li \emph{et al} define the interval $\Delta 
T^{*}=T_2^*-T_1^*$ as the width at half-height of the differential 
curve \cite{liprl}. One can approximate the peaked curve by the rectangle with sides 
at $T_1^*$ and $T_2^*$ and the height $|f'_D(T)|_{T_D}$ in such a way, 
that $1=\int_0^\infty f'_N(T)dT \approx |f'_D(T)|_{T_D}(T_2^*-T_1^*)$. 
With the account of Eq.~\eqref{dtdef} this leads to the obvious $\Delta 
T^{*}=\Delta T$, proving that both definitions of transition interval 
are equivalent, at least in the sense of asymptotic, size scaling 
relations. The same Eq.~\eqref{dtdef}, when inserted into the cooperativity 
measure Eq.~\eqref{cmdef} simply results in
\begin{equation} \label{cm1}
\Omega_c=\left( {\frac{T_D}{\Delta T}}\right)^2.
\end{equation}
\noindent The result is not surprising, since the $\frac{T_D}{\Delta T}$ ratio is 
common in the studies of finite size effects at phase transitions 
\cite{fisherprb,binder,hansmann}. In view of Eq.~\eqref{tfdtN}, valid 
for the two-state model it simply means that 
\begin{equation} \label{cm2}
\Omega_c=(\log g_D/4)^2\propto N^2.
\end{equation} 
However, if not bound to the two-state paradigm, the more 
general and model-independent formula expressed with Eq.~\eqref{cm1} 
allows to establish direct links between the well-known size scaling 
relations and the cooperativity measure $\Omega_c$. To take 
into account the possibility for both Ist and IInd order mechanisms of 
the phase transition, $\frac{T_D}{\Delta T} \propto N^{1/d\nu}$ scaling 
should be considered \cite{fisherprb,binder,hansmann,degennes}
(instead of $N^1$, used by Li \emph{et al}), where $d\nu$ is a 
critical exponent of correlation length or radius of gyration; $d\nu=1$ 
and $d\nu=2$ values would correspond to the first and the second order 
phase transition, accordingly. From Eq.~\eqref{cm1} it 
immediately follows, that 
\begin{equation} \label{cm3}
\Omega_c \propto N^{2/d\nu}.
\end{equation}
In Fig.~\ref{fig:epsart} we 
have replotted data from Ref.~\onlinecite{liprl} and compared them with 
Eq.~\eqref{cm1}. The data points for $\ln{\Omega_c}$ and 
$2\ln{\frac{T_D}{\Delta T}}$ vs $\ln{N}$ almost superimpose, and the 
corresponding fitted straight lines are indistinguishable, thus 
validating Eq.~\eqref{cm1} over the set of data from Ref.~\cite{liprl}. The 
fit resulted in $d\nu_{exp}=0.92$, which is close to, but not equal to  
one. The scaling on the basis of Eq.~\eqref{cm1} nicely fits experimental 
trends and thus allows us to treat protein folding as a true phase 
transition in a finite system in the sense of 
Lifshits-Grosberg-Khokhlov \cite{rmp1978}. The fact that the 
transition interval has the same size-scaling exponent as the 
correlation length is a nice example of the contribution of 
correlations in protein conformations to folding 
cooperativity. 

There is another experimental evidence that supports our view. Ptitsyn 
and Uversky have proposed the molten globule as the third thermodynamic 
state of protein molecules in a number of publications 
\cite{uversky-ptitsyn,ptitsyn2}. Based on the systematic analysis of 
data on urea and guanidinium chloride induced transition of globular 
proteins from the native to the unfolded state ($N \rightarrow U$), 
from the native to the molten globule ($N \rightarrow MG$) state and 
from the molten globule to the unfolded state ($MG \rightarrow U$), it 
has been shown that in all these cases the cooperativity of unfolding 
increases linearly with the increase in molecular weight of the protein 
up to $25 - 30$ kDa \cite{uversky-ptitsyn,ptitsyn2}. In fact, this 
cooperativity of all three transitions measured in terms of $\Delta n$ 
(see Eq.~\eqref{dndef}) follows the $\log \Delta n=d\nu\log M-b$, with 
$d\nu_{N-U}=0.97$, $d\nu_{N-MG}=1.02$ and $d\nu_{MG-U}=0.89$, all close to the 
$d\nu=0.92$ value, estimated from temperature inspired set of data from 
Ref.~\onlinecite{liprl}. It means, that such a dependence of the 
cooperativity of urea-induced and guanidinium chloride-induced 
transitions in small proteins on their molecular weight suggests that 
all three types of transitions are all-or-none, indicating that the 
molten globule state is separated from the native and unfolded state by 
all-or-none transitions \cite{uversky-ptitsyn,ptitsyn2}. Thus the 
experimental data on denaturant-induced unfolding of small globular 
proteins are consistent with the linear $\log \Omega_c$ vs. $\log N$ 
dependence described in Ref.~\onlinecite{liprl}.

\textbf{The comparison of cooperativity measures} shows that each of 
them has its advantages and drawbacks. The strict two-state assumption, 
expressed in Eq.~\eqref{dos} allows the derivation of $k_2 \approx 1$ 
at large $N$, which is therefore a necessary condition for the 
two-state folding. Independent of the chain length, $k_2$ allows the 
statement which of the proteins under consideration comes closer to the 
ideal two-state behaviour. Instead, in the same $N\rightarrow \infty$ 
limit, $\Omega_c$ tends to infinity, which means that under other equal 
conditions, longer chains have higher values of the cooperativity 
measure $\Omega_c$. On the other hand, $k_2$ as defined by 
Eq.~\eqref{k2}, contains both equilibrium and kinetic quantities, which 
are only equal when the system has reached equilibrium and the 
deviation from the unity can be attributed to kinetic traps (see also 
Ref.~\cite{enzym2004} for the definition and discussion about the \textit{kinetic 
cooperativity}). Regarding the $\Omega_c$, once expressed through the 
$\frac{T_D}{\Delta T}$ it becomes a criterion similar to those 
introduced in other areas of Physics to deal with the effects of finite 
size at phase transitions. The last fact puts it on very solid trails.

In summary, we have contributed to a better understanding of the 
physical basis of the two cooperation criteria under consideration. For 
the first time the size scaling expressions for the cooperation 
criteria are derived and analysed (Eqs.~\eqref{k2fin},\eqref{cm1}). As 
a result, we concluded that $k_2$ can be conveniently used to compare 
cooperativity for individual proteins, while $\Omega_c$ is more useful 
for comparing protein folding data sets with respect to size scaling analysis.

\begin{acknowledgments}
Discussions with prof. Hue Sun Chan from University of Toronto are thankfully acknowledged. A.B. acknowledges the partial financial support from Erasmus+
Project No. (2018-1-SI01-KA107-046966); A.B. and J.G. acknowledge the partial financial support from the Javna
Agencija za Raziskovalno Dejavnost RS through Project No.(J1-1705); A.B. and M.V. acknowledge the partial financial support from the Javna
Agencija za Raziskovalno Dejavnost RS through Program No.(P2-0412).
\end{acknowledgments}


\end{document}